\documentclass[pss,fleqn,embeddedheads]{w-art}

\usepackage{times}
\usepackage{w-thm}
\usepackage{amssymb}

\usepackage[]{graphicx}
\begin{document}

\DOIsuffix{theDOIsuffix}
\Volume{XX}
\Issue{1}
\Copyrightissue{01}
\Month{06}
\Year{2005}
\pagespan{1}{}
\Receiveddate{\sf zzz} \Reviseddate{\sf zzz} \Accepteddate{\sf
zzz} \Dateposted{\sf zzz}




\title{Estimation of gloss from rough surface parameters}

\author{Ingve Simonsen\footnote{Corresponding
     author: e-mail: {\sf Ingve.Simonsen@phys.ntnu.no}}\inst{1}}
\author{{\AA}ge G.\  Larsen\inst{2}} 
\author{Erik Andreassen\inst{2}}
\author{Espen Ommundsen\inst{3}}
\author{Katrin Nord-Varhaug\inst{3}}

\address[\inst{1}]{Department of physics, NTNU, NO-7491 Trondheim, Norway}
\address[\inst{2}]{SINTEF Materials and Chemistry, P.O. Box 124 Blindern, NO-0314 Oslo, Norway}
\address[\inst{3}]{Borealis AS, NO-3960 Stathelle, Norway}

\begin{abstract}
  Gloss is a quantity used in the optical industry to quantify and
  categorize materials according to how well they scatter light
  specularly. With the aid of phase perturbation theory, we derive an
  approximate expression for this quantity for a one-dimensional
  randomly rough surface.  It is demonstrated that gloss depends in an
  exponential way on two dimensionless quantities that are associated
  with the surface randomness: the root-mean-square roughness times
  the perpendicular momentum transfer for the specular direction, and
  a correlation function dependent factor times a lateral momentum
  variable associated with the collection angle.  Rigorous Monte Carlo
  simulations are used to access the quality of this approximation,
  and good agreement is observed over large regions of parameter
  space.
\end{abstract}
\maketitle                   

\renewcommand{\leftmark}
{I. Simonsen et al.: Estimation of gloss from rough surface parameters}


\section{Introduction }
\label{Sec:Intro}

Naturally occurring surfaces are not fully planar. They often show
some degree of roughness at the scale of optical
wavelengths~\cite{Beckmann,BassFuks,Ogilvy,Chew2001,SimonsenReview}. This
causes light incident upon them  to be partly reflected away
from the specular direction. The concept of gloss is related to the
amount of light scattered into a small angular interval about the
specular direction.  When designing and manufacturing materials for
which gloss is considered an essential parameter, it is desirable to
know how this quantity is related to the surface topography, and in
particular, to the parameters used to characterize them.

Gloss does in principle depend on any process that can scatter light
away from the specular direction; {\it e.g.} surface and bulk
randomness. In the present work, we will limit ourselves to situations
where the bulk randomness can be ignored relative to the surface
randomness. The gloss for randomly rough surfaces was recently studied
experimentally~\cite{Mendez,Wang} and theoretically~\cite{Barrera}
(see also Ref.~\cite{Porteus}).  In this latter study, Alexander-Katz
and Barrera derived, within the scalar Kirchhoff approximation, an
expression for gloss for two types of surface height-height
correlations (exponential and Gaussian).  These authors stressed that
it is important to {\em not} neglect the incoherent contribution to
gloss (stemming from the diffusely scattered light), since it may be
significant.  The publications~\cite{Barrera,Porteus} are among the
few studies found in the literature where also the surface correlation
is included when trying to estimate gloss from the surface roughness
parameters.

In this paper, we reexamine gloss of a randomly rough surface. An
expression for this industrially relevant quantity is derived within
the framework of phase-perturbation theory~\cite{Shen,SG}, for a
general height correlation function. Like the authors of
Ref.~\cite{Barrera}, it is found that the surface correlation is
important for gloss in general.  However, contrary to Alexander-Katz
and Barrera~\cite{Barrera}, it turns out that in our formulation it is
not the product of the correlation length and a lateral momentum
variable (collection angle) that acts as one of the reduced variables
for the estimation of gloss from surface parameters.  Instead, one
finds that the corresponding reduced variable depends on the
{\em functional} form of the correlation function satisfied by the surface
roughness, and not only its correlation length.  Only in some limiting
cases do we recover results consistent with Ref.~\cite{Barrera}.

When applying approximate expressions to calculate important optical
quantities, it is of utmost importance to be able to have knowledge
about the range of validity of the expression used. In order to
address this crucial point, we use results from rigorous Monte Carlo
simulations to gauge the quality of the proposed approximate
analytical expression to gloss. To the best of out knowledge, such
extensive comparison to rigorous simulation results has never been
reported before. The main reason for considering a one-dimensional
scattering geometry was to enable such a comparison. Rigorous
simulation in this context of a two-dimensional geometry is out of
reach for todays computer power.


\section{Gloss}
\label{Sec:Gloss}

In the optical industry, {\em gloss} is used extensively to quantify
the visual appearance and functional properties of various
materials~\cite{Mendez,Barrera,NIST,GlossStandards}.
However, it does not show the same prominence in the optics branch of
science where it is not-so-often considered. In fact, the term does
not even have a rigorous scientific definition, and it comes in
several ``flavors'': specular and wide-angle gloss referring to
different angles of incidence ($20^\circ$, $60^\circ$ and
$85^\circ$)~\cite{GlossStandards}.  For instance, the way that gloss
is quantified and measured (by gloss-meters) depends on industrial
standards~\cite{GlossStandards} that are different in, say,
north-America and Europe.  However, in essence, what the gloss numbers
quantify, is how well a material scatter light incident upon it into
the specular direction. Depending on the level of surface topography
and/or inhomogeneties in the bulk of the material, a fraction of the
incident light will be scattered into directions other than the
specular. Such mechanisms will contribute to reduce the gloss numbers
of such materials.

In this study, we will define gloss, ${\cal G}(\theta_0)$, as the
fraction of the incident light that is scattered into a small angular
interval, $\Delta \theta$, about the specular
direction.\footnote{Usually the normalization is {\em not} done with
  respect to the total scattered energy, but instead relative to the
  (specular) reflectance of a smooth reference (black glass) material
  (that depends on the standard used). The authors of
  Ref.~\cite{Barrera} adopted a normalization consisting of the
  reflectance of a {\em smooth} material of the {\em same} type as
  that being investigated. For not too rough surfaces, our definition
  follows closely the one by Alexander-Katz and Barrera since the
  total scattered energy is only moderately sensitive to the
  roughness.}  This definition is not identical to any of the
industrial standards in common use today~\cite{GlossStandards}.
However, it shares the main characteristics of these standards, and
for appropriate choices of the angular interval can be related to
them. Furthermore, we will be somewhat unorthodox and also consider
``gloss'' in transmission, even though it is not a commonly used term.
The concept of {\em haze} is more customarily considered in this
context~\cite{HazeGloss}.


\section{The Scattering Geometry}
\label{Sec:Geometry}

\begin{figure}[t]
  \centering
  \includegraphics[width=7cm,height=5cm]{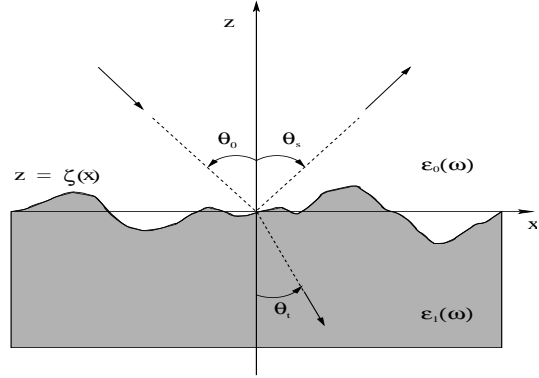}
\caption{The scattering geometry used in this
  study. The rough surface is defined by $z=\zeta(x)$. The region
  above the surface, $z>\zeta(x)$, is assumed to be vacuum
  ($\varepsilon_0(\omega)=1)$, while the medium below is a dielectric
  characterized by a frequency-dependent dielectric function
  $\varepsilon_1(\omega)$. Notice for which directions the angle of
  incident ($\theta_0$), scattering ($\theta_s$), and transmission
  ($\theta_t$) are being defined as positive. An angle of transmission
  is only well-defined if the lower medium is transparent, {\it i.e.} 
  if $\Re e\, \varepsilon_1(\omega)>0$.  }
    \label{fig:Geometry} 
\end{figure}

The scattering geometry that we will consider in this study is
depicted in Fig.~\ref{fig:Geometry}. In the region $z > \zeta(x)$ it
consists of vacuum ($\varepsilon_{0}(\omega)=1$) and for $z <
\zeta(x)$ of a dielectric medium characterized by an isotropic,
frequency-dependent, dielectric function $\varepsilon_{1}(\omega)$.
Here $\zeta(x)$ denotes the surface profile function, that is assumed to
be a single-valued function of $x$ that is differential as many times
as is necessary.  Furthermore, it constitutes a zero mean, stationary,
Gaussian random process that is defined by 
\begin{subequations}
  \label{eq:surf-roughness}
\begin{eqnarray}
  \left< \zeta(x)\right> &=& 0, \\
  \left< \zeta(x)\zeta (x')\right> &=& \sigma^2 W(|x-x'|). 
\end{eqnarray}
\end{subequations}
Here $W(|x|)$ denotes the (normalized) auto-, or height-height
correlation function, $\sigma$ is the root-mean-square of the surface
roughness, and $\left<\cdot\right>$ denotes the average over an
ensemble of realizations of the surface roughness. For the discussion
to follow one will also need the power spectrum of the surface
roughness, defined as the Fourier transform of the correlation
function, {\it i.e.}
\begin{eqnarray}
  \label{eq:power-spec}
  g(|k|) &=& \int dx\, e^{-ikx}W(|x|). 
\end{eqnarray}

\section{An approximate expression to gloss}
\label{Sec:Approximation}

The mean differential reflection and transmission coefficients,
collectively denoted $\left< \partial U/\partial\theta \right>$, are
two experimentally and theoretically accessible quantities frequently
used to study the angular distribution of the reflected or transmitted
light~\cite{SimonsenReview}. They express the fraction of the power
incident upon the surface that is scattered (or transmitted) into an
angular interval $d\theta$ about the angle $\theta_s$ (or $\theta_t$). 
Hence, gloss as defined in Sec.~\ref{Sec:Gloss}, can mathematically
be defined according to
\begin{eqnarray}
  \label{eq:gloss-int}
  {\cal  G}(\theta_0)   &=&  
  \frac{1}{{\cal U}} \;
  \int^{\theta_+}_{\theta_-}d\theta \, 
  \left<\frac{\partial U}{\partial\theta}\right>,
\end{eqnarray}
where $\theta_\pm=\Theta\pm\Delta\theta$, with $\Theta = \arcsin
\left\{ (\sqrt{\varepsilon_m}/\sqrt{\varepsilon_0}) \sin\theta_0
\right\}$ being the specular direction in reflection
($\varepsilon_m\!=\!\varepsilon_0$) and transmission
($\varepsilon_m\!=\!\varepsilon_1$).  Moreover, ${\cal U} =
\int^{\pi/2}_{-\pi/2}\,d\theta\left<\partial U/\partial\theta\right>$
denotes the reflectance or transmittance of the rough surface.  Within
the framework of phase-perturbation theory~\cite{Shen,SG,HazeGloss},
it can be demonstrated that gloss of a randomly rough surface can be
expressed as~\cite{HazeGloss}
\begin{subequations}
  \label{eq:gloss-final-all}
  \begin{eqnarray}
    \label{eq:gloss-final}
    {\cal G}(\theta_0)
    &\simeq& 
    \exp\left[-\sigma^2\Lambda^2(k|k)
      \left(1-\frac{G(a) \Delta q}{\pi} \right)
    \right], 
       \qquad \Lambda(q|k) = \left\{
         \begin{array}{cl}
           \alpha_0(q)+\alpha_0(k) & \mbox{Refl.} \\
           \alpha_1(q)-\alpha_0(k) & \mbox{Trans.} \\
         \end{array} \qquad
         \right.
  \end{eqnarray} 
  where $k=\sqrt{\varepsilon_0}(\omega/c)\sin\theta_0$ and
  $\alpha_m(q) = \sqrt{\varepsilon_m\omega^2/c^2-q^2}$
  ($\Im m\,\alpha_m>0$). Moreover, in writing Eq.~(\ref{eq:gloss-final}),
  one has introduced a power spectrum dependent factor defined
  according to\footnote{For an exponential and Gaussian correlation
    function characterized by the correlation length $a$, one has
    $G(a) = (2/\Delta q)\arctan(\Delta q a)$ and $G(a)=(\pi/\Delta q)
    \, \mbox{erf}(\Delta q a/2)$, respectively, where
    $\mbox{erf}(\cdot)$ denotes the error function.}
  \begin{eqnarray}
    \label{eq:APS}
    G(a) &=& \frac{1}{2\Delta q}
    \int^{q_+}_{q_-} dq \, g(|q-k|)
    \;=\;
    \frac{1}{2\Delta q}
    \int^{\Delta q}_{-\Delta q}dq \, g(|q|),
  \end{eqnarray}
  where $q_\pm=k\pm\Delta
  q=\sqrt{\varepsilon_m}(\omega/c)\sin\theta_\pm$ with 
  $\Delta q=\sqrt{\varepsilon_m}(\omega/c)\cos\Theta\sin\Delta\theta$,
  and $a$ denotes the correlation length.
\end{subequations}

There are several important observations that should be made from
Eqs.~(\ref{eq:gloss-final-all}). First, gloss can be expressed in
terms of {\em two dimensionless quantities}: $\sigma\Lambda(k|k)$ and
$G(a) \Delta q$.  The former is the product of the root-mean-square
roughness associated with the surface topography and the perpendicular
momentum transfer of the scattering (or transmission) process into the
specular direction. Hence, it does depend on the ``amount'' of
roughness {\em but not} on how it is being correlated along the
interface. In quite a few studies of gloss for rough surfaces, this is
the only factor considered.  The latter quantity, $G(a) \Delta
q$, on the other hand, depends on the form of the power spectrum and
thus, indirectly, on the correlation length. Notice, however, that it
is only the functional form of the power spectrum $g(\left|q\right|)$
around zero (lateral) momentum transfer that enters into the
expression for gloss via $G(a)$.  Second, the dependence of
gloss on the angle of incidence {\em only} enters through the
perpendicular momentum transfer present in $\sigma\Lambda(k|k)$
(with $k=\sqrt{\varepsilon_0}(\omega/c)\sin\theta_0$).
 
For completeness, and to facilitate the use of these approximate
expressions for gloss, we also give the full expressions in terms of
the ``defining'' quantities. Gloss in reflection can be expressed as
\begin{subequations}
  \begin{eqnarray}
    \label{eq:haze-ref}
     {\cal G}_s(\theta_0) &\simeq&  
       \exp\left[
            -16\pi^2 \varepsilon_0 \left(\frac{\sigma}{\lambda}\right)^2\cos^2\theta_0\,  
            \left\{1- 2\sqrt{\varepsilon_0}
                   \frac{G(a)}{\lambda}
                   \sin\Delta\theta   
                  \cos\theta_0 
                 \right\}
            \right],
  \end{eqnarray}
  while in transmission one has  
  \begin{eqnarray}
    \label{eq:haze-trans}
     {\cal G}_t(\theta_0) &\simeq&  
       \exp\left[
     -4\pi^2 \varepsilon_0 \left( \frac{\sigma}{\lambda}\right)^2
  \left\{\sqrt{\frac{\varepsilon_1}{\varepsilon_0} -\sin^2\theta_0}
        -\cos\theta_0 \right\}^2 
          \right.\nonumber \\ && \left. \qquad \quad \times 
  \left\{1 - 2\sqrt{\varepsilon_1}
                   \frac{G(a)}{\lambda}
                   \sin\Delta\theta   
                   \sqrt{1-\frac{\varepsilon_0}{\varepsilon_1}\sin^2\theta_0}
             \right\}
        \right]. 
  \end{eqnarray}
\end{subequations}
In obtaining the expression for ${\cal G}_t(\theta_0)$ it has been
used that
$\cos\Theta_t=\sqrt{1-(\varepsilon_0/\varepsilon_1)\sin^2\theta_0}$
for the specular direction in transmission.  

Previously, Alexander-Katz and Barrera~\cite{Barrera}, reported while
using scalar Kirchhoff theory, that the reduced variables for gloss
(in reflection) were $(\sigma/\lambda)\cos\theta_0$ and
$(a/\lambda)\Delta\theta$.  Since in the radiative
region~\cite{Chew2001,SimonsenReview}
$\alpha_0(k)=\sqrt{\varepsilon_0}(\omega/c)\cos\theta_0$, it follows
readily that $\sigma\Lambda(k|k)\propto (\sigma/\lambda)\cos\theta_0$.
However, we do not in general find that $G(a)\Delta q$ scales like
$a\Delta \theta$ (the product of the correlation length and the
collection angle). Only in the limit $a\Delta q \ll 1$, for which
$G(a)\propto a$, do we recover the scaling reported by Alexander-Katz
and Barrera~\cite{Barrera}. Notice that for small collection angles
one has $a\Delta q\simeq 2\pi\sqrt{\varepsilon_m} \cos\Theta
\,(a/\lambda)\Delta\theta$.  However, below we will see that when the
ratio $a/\lambda$ becomes rather small, the phase perturbative
approximation to gloss becomes less accurate. Over the range of
validity of this approximation we therefore find that for a
one-dimensional roughness the reduced correlation dependent variable
for gloss is $G(a)\Delta q$ and not simply the product of the
correlation length and the collection angle as reported in
Ref.~\cite{Barrera}.

\begin{figure}[t]
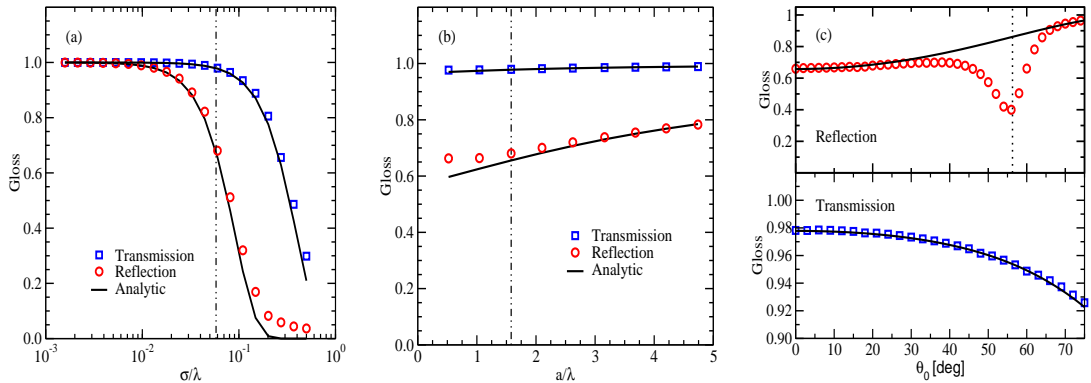

  \centering
  \includegraphics*[width=4.5cm,height=5cm]{gloss_vs_rms-100}   \quad
  \includegraphics*[width=4.5cm,height=5cm]{gloss_vs_corr-037}  \quad
  \includegraphics*[width=4.5cm,height=5cm]{gloss_vs_angle-100-037}
  \caption{Gloss as a function of (a) the surface roughness
    $\sigma/\lambda$ for fixed $a/\lambda=1.58$ and
    $\theta_0=0^\circ$; (b) the correlation length $a/\lambda$ for
    $\sigma/\lambda=0.058$ and $\theta_0=0^\circ$; (c) the angle of
    incidence $\theta_0$ keeping $\sigma/\lambda=0.058$ and
    $a/\lambda=1.58$ fixed.  The surface roughness used to obtain
    these results was a Gaussian random process characterized by an
    exponential correlation function, $W(x) = \exp(-|x|/a)$, of
    correlation length $a$.  For all figures the wavelength of the
    $p$-polarized incident light was $\lambda=0.6328 \, \mu m$ and
    $\Delta \theta=2.5^\circ$. The open symbols are results of
    rigorous Monte Carlo simulations, while the solid lines are the
    predictions of Eqs.~(\protect\ref{eq:gloss-final-all}). The
    vertical dashed-dotted lines in
    Figs.~\protect\ref{fig:Gloss-scaling}(a) and (b) correspond to the
    assumptions made for $\sigma/\lambda$ and $a/\lambda$ in
    Figs.~\protect\ref{fig:Gloss-scaling}(b) and (a), respectively.
    Moreover, the vertical dash line in
    Fig.~\protect\ref{fig:Gloss-scaling}(c) corresponds to the
    position of the Brewster angle for the corresponding planar
    scattering geometry. Recall that in $s$-polarization the Brewster
    phenomenon is not present, and the prediction of
    Eqs.~(\protect\ref{eq:gloss-final-all}) is of good quality for
    the entire angle of incidence range. }
    \label{fig:Gloss-scaling} 
\end{figure}

We will now address the accuracy of the analytic expressions for gloss
(\ref{eq:gloss-final-all}) derived in this work. This will be achieved
by comparing these expressions to what can be obtained from rigorous
Monte Carlo simulations~\cite{Chew2001,SimonsenReview} that in principle takes
{\em all} higher order scattering processes into account.  The
simulation results using an exponential correlation function (and
assuming $p$-polarization for the incident light) are presented in
Fig.~\ref{fig:Gloss-scaling}.  They demonstrate that the approximate
expressions are rather good even for relatively rough surfaces.
Furthermore, they seem to produce the best results for
$a/\lambda\gtrsim 1$ (and $\sigma/\lambda$ not too large) for which
one naively would expect single scattering to mainly contribute.
Hence, for rough, shortly correlated surfaces where $\sigma/a\gg 1$
the approximate expressions presented herein are no longer adequate.
Comparable, or better, results have been obtained for
$s$-polarization. There is one important difference to be noticed
between the results for $p$ and $s$-polarization. In the former state
of polarization, the Brewster angle phenomenon is present, and as a
result causes the estimation of gloss to be rather
sensitive\footnote{The obtained results for gloss do also for such
  angles of incidence depend somewhat on the definition used for
  gloss.} to angles of incidence about the Brewster angle (cf.
Fig.~\ref{fig:Gloss-scaling}(c)). However, for $s$-polarization,
Eqs.~(\protect\ref{eq:gloss-final-all}) represent a good approximation
to gloss over the entire range of angles of incidence.


\section {Conclusions}

In conclusion, we have derived approximate analytic expressions to
gloss within the framework of phase-perturbation theory.  We found
that the reduced variables for gloss are: ({\it i}) the
root-mean-square roughness times the perpendicular momentum transfer
for the specular direction, and ({\it ii}) a height-height correlation
dependent factor times a lateral momentum transfer variable. These
findings only partly agree with previous reported results.  The
precision of the analytic expressions to gloss in terms of parameters
normally used to characterize randomly rough surfaces was gauged by
comparison to Monte Carlo simulations.  Good agreement was found over
large regions of parameter space, also for rather rough surfaces. In
particular when the correlation length was of the order of, or larger
than, the wavelength, the agreement was excellent.




\begin{thebibliography}{10}


 \bibitem{Beckmann} P.\ Beckmann and A.\ Spizzichino, {\sl The
 scattering from electromagnetic waves from rough surfaces}, (Artech
 House, 1963). 

 \bibitem{BassFuks}
   F.G. Bass and I.M. Fuks, {\sl Wave scattering from statistically
     rough surfaces}, (Pergamon Press, Oxford, UK, 1979). 

 \bibitem{Ogilvy} 
   J.\ A.\ Ogilvy, {\sl Theory of wave scattering from
     random rough surfaces} (IOP Pub., Bristol, 1991). 

\bibitem{Chew2001}
  K.F. Warnick and W.C. Chew, 
  Waves Random Media {\bf 11}, R1-R30 (2001).

\bibitem{SimonsenReview}
  I.\ Simonsen, {\em A random walk through surface scattering phenomena:
    Theory and phenomenology},  arXiv:cond-mat/0408017, 2004.

  
 \bibitem{Mendez}
   R.\ R.\ M\'endez, R.\ G.\ Barrera, and R.\ Alexander-Katz,
  Physica A {\bf 207}, 137 (1994). 

 \bibitem{Wang}
   L. Wang, T. Huang, M.R. Kamal, A.D. Rey, and J. Teh,
   Polymer Engineering \& Science {\bf 40}, 747 (2004).

 \bibitem{Barrera} 
   R.\ Alexander-Katz and R.\ G.\ Barrera, %
   J. Polym. Sci. B: Polym. Phys. {\bf 36}, 1321 (1998). 

\bibitem{Porteus}
  J.\ O.\ Porteus, J. Opt. Soc. Am. {\bf 53}, 1394 (1964).

\bibitem{Shen}
   J.\ Shen and A.\ A.\ Maradudin,
    Phys.\ Rev.\ B {\bf 22}, 4234 (1980). 

\bibitem{SG}
  J.\ A.\ S\'anchez-Gil, A.\ A.\ Maradudin, and E.\ R.\ M\'endez,
  J. Opt. Soc. Am. A {\bf 12}, 1547 (1995). 


\bibitem{NIST}
 M.E. Nadal and E.A. Thompson, 
Journal of Coatings Technology {\bf 72}, 61 (2000). 

\bibitem{GlossStandards}
 ASTM Standard D 523-89, Standard Test Method for Specular Gloss,
 1989; ASTM Standard E 430-91, Standard Test Method for Measurement
 of Gloss of High-Gloss Surfaces by Goniophotometry, 1991.  
 
\bibitem{HazeGloss}
  Ingve Simonsen {\it et al.} unpublished 2005.

 




























  















 









\end{thebibliography}
\end{document}